# A NEW S-S' PAIR CREATION RATE EXPRESSION IMPROVING UPON ZENER CURVES FOR I-E PLOTS

*A. W. Beckwith*

*Department of Physics and Texas Center for Superconductivity and Advanced Materials at the University of Houston*

*Houston, Texas 77204-5005 USA*

**ABSTRACT**

To simplify phenomenology modeling used for charge density wave (CDW) transport, we apply a wavefunctional formulation of tunneling Hamiltonians to a physical transport problem characterized by a perturbed washboard potential. To do so, we consider tunneling between states that are wavefunctionals of a scalar quantum field $\phi$. ***I-E*** curves that match Zener curves — used to fit data experimentally with wavefunctionals congruent with the false vacuum hypothesis. This has a very strong convergence with the slope of graphs of electron-positron pair production representations. The newly derived results include a threshold electric field explicitly as a starting point without an arbitrary cut off value for the start of the graphed results, thereby improving on both the Zener plots and Lin's generalization of Schwingers 1950 electron-positron nucleation values results for low dimensional systems. The similarities in plot behavior of the current values after the threshold electric field values argue in favor of the Bardeen pinning gap paradigm proposed for quasi-one-dimensional metallic transport problems.

Correspondence: A. W. Beckwith: **projectbeckwith2@yahoo.com**.

**I. INTRODUCTION**

As of 1985, Dr. John Miller used a Zener curve fitting polynomial to match qualitatively current vs. electric field values data he obtained from applying an electric field to a $NbSe_3$ crystal at low temperatures.[1] What is presented in this article is a derivation of an analytical current expression that improves upon the Zener curve fit used by Dr. Miller et al. and uses a pinning gap presentation of tunneling as an alternative to Gruner's classical model of charge density wave (CDW) transport.[2] This new derivation permits an improvement upon Schwinger's[3] work in the 1950s predicting the likelihood of formation of electron-positron pairs. After a threshold electric field value the *I-E* plots due to this new theoretical construction are a good qualitative fit once when Lin[4] generalized Schwigner's[3] results for general dimensions are viewed at the lowest permitted dimensionality possible.

In short, we argue that a qualitatively good fit between our analytical expression and a multi-dimensional generalization of Schwinger's[3] expression, which in Lin[4] was restricted to the one-dimensional case for electron-positron pair nucleation, shows that a pinning gap interpretation of tunneling in quasi-one-dimensional systems for charge density waves (CDW) is appropriate and optimal for experimental data sets. This supports the conclusion that our analytical results improvement over the earlier Zener curve fit approximation given by Miller et al[1] was not accidental and contains serious physics worthy of consideration.

Herein, we will present a new CDW transport physics procedure for calculating current *I* vs. *E* electric field plots for quasi-one-dimensional metals, assuming:

i. The current *I* is directly proportional to the modulus of the diagonal terms for the tunneling Hamiltonian. The reasons for this non-standard choice will become clear later on in the text. This tunneling Hamiltonian uses a functional integral version of an expression used initially by Tinkham[5] for scanning electron microscopy.

ii. Gaussian wavefunctionals are chosen to represent the initial and final states of a soliton-anti soliton pair (S-S') traversing a pinning gap presentation of impurities in a quasi-one-dimensional metallic lattice. These wavefunctionals replace the wave functions Tinkham[5] used in his tunneling Hamiltonian (T.H.) matrix element and are real-valued.

iii. The pinning gap means that a washboard potential with small driving term[6] $\mu_E \cdot (\phi - \Theta)^2$ added to the main potential term of the washboard potential, is used to model transport phenomenology. We also argue that this potential permits domain wall modeling of S-S' pairs.[7] In this situation, $\mu_E$ is proportional to the electrostatic energy between the S-S' pair constituents (assuming a parallel plate capacitor analogy); $\Theta$ is a small driving force we will explain later, dependent upon a ratio of an applied electric field over a threshold field value. As we show later, the dominant washboard potential term will have the value of (pinning energy) times $(1 - \cos \phi)$

It is useful to note that Kazumi Maki,[8] in 1977, gave the first generalization of Sidney Coleman's least action arguments[9] to NbSe$_3$ electrodynamics. We use much the same pinning potential, with an additional term due to capacitance approximation of energy added by the interaction of a S-S' pair with each other.[6] While Dr. Maki's work is very complete, it does not include in a feature we found of paramount importance, that of the effects of a threshold electric field value to 'turn on' effective initiation of S-S' pair transport across a pinning gap. It is also relevant to note that we previously found[10] that topological soliton style arguments can explain why the potential lead to the least action integrand collapsed to primarily a quadratic potential contribution, which permits treating the wave functional as a Gaussian. As would be expected, the ratio of the coefficient of pinning gap energy of the Washboard potential used in $NbSe_3$ modeling to the quadratic term $\mu_E \cdot (\phi - \Theta)^2$ used in modeling energy stored in between S-S' pairs was fixed by experiment to be nearly 100 to 1 , which is a datum we used in our calculations.[11]

In several dimensions, we find that the Gaussian wavefunctionals would be given in the form given by Lu.[12] Lu's integration given below is a two dimensional Gaussian wave functional. The analytical result we are working with is a one-dimensional version of a ground-state wave functional of the form[12]

$$|0>^o = N \cdot \exp\left\{-\int_{x,y}\left[(\phi_x - \varphi) \cdot f_{xy} \cdot (\phi_y - \varphi)\right] \cdot dx \cdot dy\right\} \tag{1}$$

Lu's Gaussian wave functional is for a non-perturbed, Hamiltonian as given in Eq. (2) below

$$H_O = \int_x \left[\frac{1}{2} \cdot \Pi_x^2 + \frac{1}{2} \cdot (\partial_x \phi_x)^2 + \frac{1}{2} \cdot \mu^2 \cdot (\phi_x - \varphi)^2 - \frac{1}{2} \cdot I_0(\mu)\right] \cdot dx \cdot dy \tag{2}$$

We should note that Lu intended the wavefunctional given in Eq. (1) to be a test functional, much as we would do for finding an initial test functional , using a simple Gaussian in computing the ground state energy of a simple Harmonic oscillator variational derivative. calculation. We may obtain a 'ground state' wave functional by taking the one dimensional version of the integrand given in Eq. (1). This means have[12] a robust Gaussian. Lu' Gaussian wave functional set t

$$\frac{\partial^2 \cdot V_E}{\partial \cdot \phi^a \cdot \partial \cdot \phi^b} \propto f_{xy} \tag{3}$$

Here, we call $V_E$ a (Euclidian-time style) potential, with subscripts $a$ and $b$ referring to dimensionality; and $\phi_x$ an 'x dimension contribution' of alternations of 'average' phase $\varphi$, as well as $\phi_y$ an 'y dimension contribution' of alternations of 'average' phase $\varphi$. This average phase is identified in the problem we are analyzing as $\phi_C$

This leads to writing the new Gaussian wavefunctional to be looking like[13,14]

$$\Psi \equiv c \cdot \exp(-\alpha \cdot \int dx [\phi - \phi_C]^2) \tag{5}$$

Making this step from Eq. (1) to Eq. (3) involves recognizing, when we go to one-dimension, that we look at a washboard potential with pinning energy contribution from $D \cdot \omega_P^2$ in one- dimensional CDW systems

$$\frac{1}{2} \cdot D \cdot \omega_P^2 \cdot (1 - \cos\phi) \approx \frac{1}{2} \cdot D \cdot \omega_P^2 \cdot \left(\frac{\phi^2}{2} - \frac{\phi^4}{24}\right) \tag{6}$$

The fourth-order phase term is relatively small, so we look instead at contributions from the quadratic term and treat the fourth order term as a small perturbing contribution to get our one dimensional CDW potential, for lowest order, to roughly look like Eq. (5). In addition, we should note that the c is due to an error functional-norming procedure, discussed below; $\alpha$ is proportional to one over the length of distance between the constituent components of a S-S' pair; the phase value, $\phi_C$, is set to represent a configuration of phase in which the system evolves to/from in the course of the S-S' pair evolution. This leads to [13,14]

$$c_1 \cdot \exp\left(-\alpha_1 \cdot \int d\tilde{x} [\phi_F]^2\right) \cong \Psi_{initial} \tag{7}$$

As well as

$$c_2 \cdot \exp\left(-\alpha_2 \cdot \int d\tilde{x} [\phi_T]^2\right) \cong \Psi_{final} \tag{8}$$

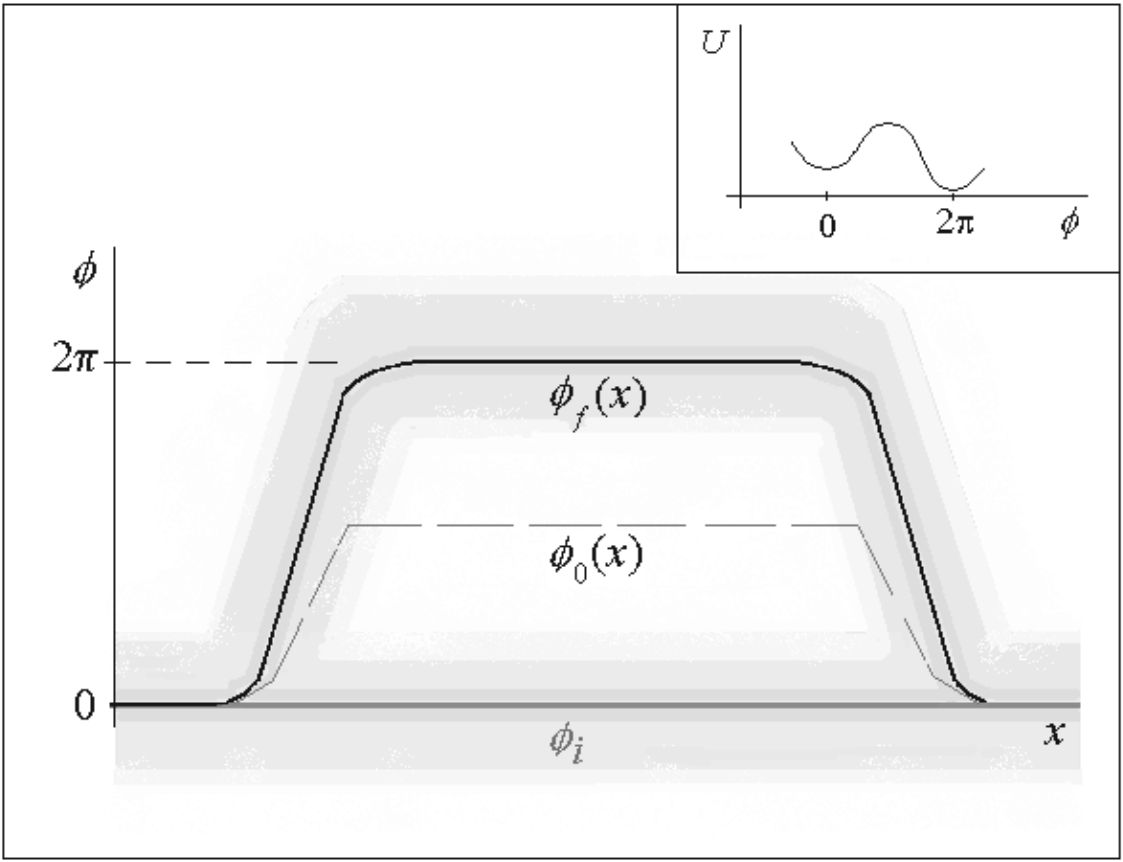

**Fig. 1:** *Evolution from an initial state $\Psi_i[\phi]$ to a final state $\Psi_f[\phi]$ for a double-well potential (inset) in a 1-D model, showing a kink-antikink pair bounding the nucleated bubble of true vacuum. The shading illustrates quantum fluctuations about the initial and final optimum configurations of the field, while $\phi_0(x)$ represents an intermediate field configuration inside the tunnel barrier. The upper right hand side of this figure is how the fate of the false vacuum hypothesis gives a difference in energy between false and true potential vacuum values*

In the current vs. applied electric field derivation results, we identify the $\Psi_i[\phi]$ as the initial wave function at the left side of a barrier and $\Psi_f[\phi]$ as the final wave function at the right side of a barrier. Note that Tekman[5] extended the tunneling Hamiltonian method to encompass more complicated geometries. We notice that when the matrix elements $T_{kq}$ are small, we calculate the current through the barrier using linear response theory. This may be used to describe coherent Josephson-like tunneling of either Cooper pairs of electrons or boson-like particles, such as super fluid He atoms.[13,14] In this case, the supercurrent is linear with the effective matrix element for transferring a pair of electrons or transferring a single boson, as shown rather elegantly in Feynman's derivation[15] of the Josephson current-phase relation. This means a current density proportional to $|T|$ rather than $|T|^2$ since tunneling, in this case, would involve coherent transfer of individual (first-order) bosons rather than pairs of fermions.[14]

Note that the initial and final wave functional states were in conjunction with a pinning gap formulation of a variation of typical band calculation structures. Fig. 2 shows much of the layout for how a tilted band structure due to an applied electric field influenced the geometry of the perturbed Washboard potential problem in the situation represented in Fig. 2:

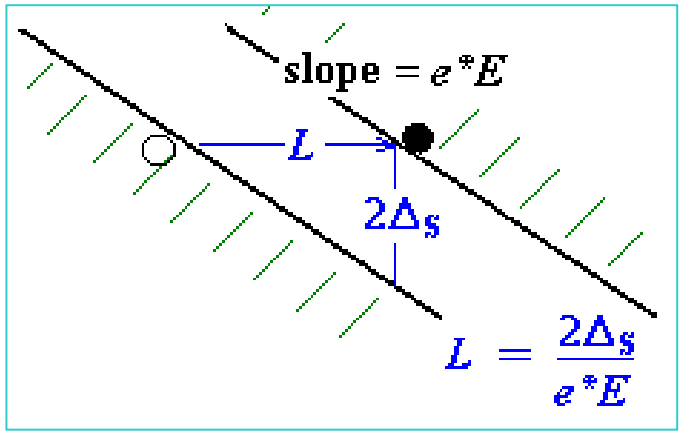

**Fig 2**: *This is a representation of 'Zener' tunneling through pinning gap with band structure tilted by applied E field*

We consider that we will be working with a Hamiltonian of the form

$$H = \int_x \left[ \frac{1}{2 \cdot D} \cdot \Pi_x^2 + \frac{1}{2} \cdot (\partial_x \phi_x)^2 + \frac{1}{2} \cdot \mu_E \cdot (\phi_x - \Theta)^2 + \frac{1}{2} \cdot D \cdot \omega_P^2 \cdot (1 - \cos \phi_x) \right] \tag{9}$$

where the potential system leads to the phenomenology represented in Fig. 2 with what we have been calling $V_E$ the Euclidian action version of the potential given above. In addition, the first term is the conjugate momentum.. Note, this conjugate momentum is our kinetic energy Specifically, we found that we had $\Pi_x \equiv D \cdot \partial_t \phi_x$ as canonical momentum density, $D \equiv \left( \dfrac{\mu \cdot h}{4 \cdot \pi \cdot v_F} \right)$, (where $\mu \equiv \dfrac{M_F}{m_{e^-}} \cong 10^3$ is a Frohlich to electron mass ratio, and $v_F$ is a Fermi velocity $> 10^3$ cm/sec), and $D \cdot \omega_P^2$ as the pinning energy. In addition, we have that $\mu_E$ is electrostatic energy, which is analogous to having a S-S' pair represented by a separation L and of cross-sectional area A, which produces an internal field [6] $E^* = \left( e^* / \varepsilon \cdot A \right)$, where $e^\bullet \cong 2 \cdot e^- \equiv$ effective charge and $\varepsilon \equiv 10^8 \cdot \varepsilon_0$ is a huge dielectric constant. Finally, the driving force term[6], $\Theta = 2 \cdot \pi \cdot \dfrac{E}{E^*}$, where the physics of the term given by $\int dx \cdot \mu_E \cdot (\phi - \Theta)^2$, leads to no instanton tunneling transitions if $\Theta < \pi \Leftrightarrow E < \dfrac{E^*}{2}$ which was the basis of a threshold field of the value $E_T = E^*/2$ due to conservation of energy considerations[6]. Finally, it is important to note that experimental constraints as noted in the device development laboratory lead to $.01 < \mu_E / D \cdot \omega_P^2 \leq .015$, which we claim has also been shown to be necessary due to topological soliton arguments.

This pinning gap structure relates to the S-S' pair formation. This can most easily be seen in the following diagram of how the S-S' pair structure arose in the first place, as given by Fig. 3:

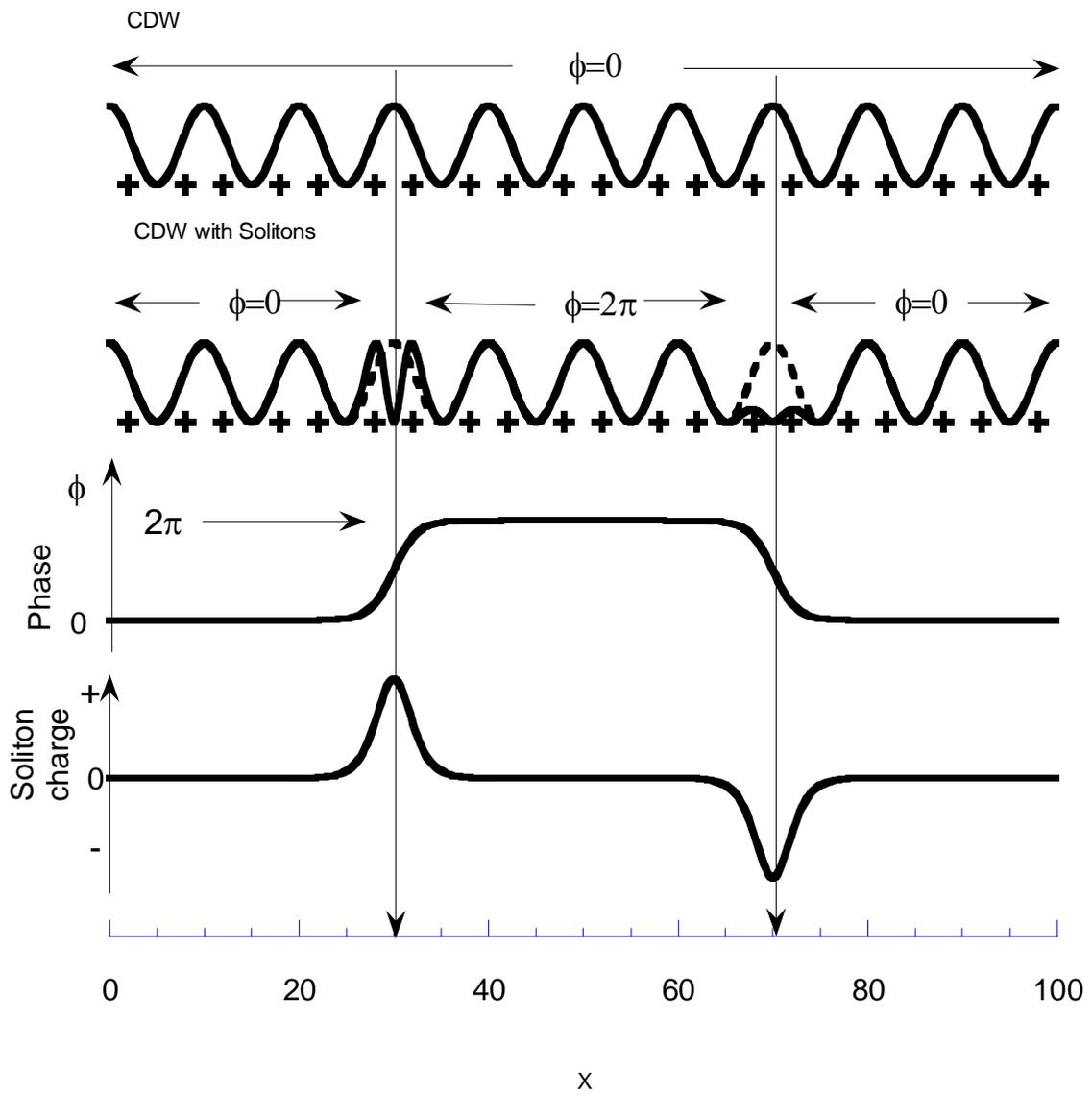

**Fig 3** : *The above figures represents the formation of soliton-anti soliton pairs along a 'chain'. The evolution of phase is spatially given by*

$$\phi(x) = \pi \cdot [\tanh b(x - x_a) + \tanh b(x_b - x)]$$

The tunneling Hamiltonian incorporates wavefunctionals whose Gaussian shape keeps much of the structure as represented by Fig. 3.

The wavefunctionals used in this problem have coefficients in front of the integrals of the phase evolutions for the initial and final states, which are the same. This meant setting the $\alpha \approx L^{-1}$ as inversely proportional to the distance between a S-S' pair.[13,14]

Following the false vacuum hypothesis,[8] we have a false vacuum phase value $\phi_F \equiv <\phi>_1 \cong$ *very small value*, as well as having in CDW, a final true vacuum[4] $\phi_T \cong \phi_{2\pi} \equiv 2\cdot\pi + \varepsilon^+$. This led to Gaussian wavefunctionals with a simplified structure. For experimental reasons, we need to have[13,14] (if we set the charge equal to unity, dimensionally speaking)

$$\alpha \approx L^{-1} \equiv \Delta E_{gap} \equiv V_E(\phi_F) - V_E(\phi_T)) \qquad (10)$$

This is equivalent to the situation as represented by Fig. 4.

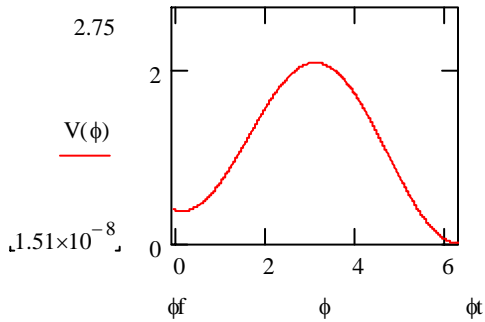

**Fig 4**: *Fate of the false vacuum representation of what happens in CDW. This shows how we have a difference in energy between false and true vacuum values*

This assumes we are using the following substitutions in the wavefunctionals

$$\Psi_f[\phi(\mathbf{x})]\Big|_{\phi \equiv \phi_{Cf}} =$$
$$c_f \cdot \exp\left\{-\int d\mathbf{x}\, \alpha\left[\phi_{Cf}(\mathbf{x}) - \phi_0(\mathbf{x})\right]^2\right\} \to \quad (11)$$
$$c_2 \cdot \exp\left(-\alpha_2 \cdot \int d\widetilde{x}\,[\phi_T]^2\right) \cong \Psi_{final},$$

and

$$\Psi_i[\phi(\mathbf{x})]\Big|_{\phi \equiv \phi_{Ci}}$$
$$= c_i \cdot \exp\left\{-\alpha \int d\mathbf{x}\,[\phi_{ci}(\mathbf{x}) - \phi_0]^2\right\} \to \quad (12).$$
$$c_1 \cdot \exp\left(-\alpha_1 \cdot \int d\widetilde{x}\,[\phi_F]^2\right) \equiv \Psi_{initial},$$

## III. EVALUATING THE TUNNELING HAMILTONIAN ITSELF TO GET A CURRENT' CALCULATION IN CDW

Our wavefunctionals plus the absolute value of the tunneling Hamiltonian in momentum space lead to, after a lengthy calculation,[13,14] a way to predict how the modulus of diagonal tunneling matrix elements that are equivalent to current will be influenced by an applied electric field. It was done in momentum space, among other things. Initially, the tunneling matrix element in momentum space had the form of

$$T_{if} \cong \frac{\hbar^2}{2\mu} \int \left( \Psi_{initial}^* \frac{\delta^2 \Psi_{final}}{\delta\phi(x)_2} - \Psi_{final} \frac{\delta^2 \Psi_{initial}^*}{\delta\phi(x)_2} \right) \vartheta(\phi(x) - \phi_0(x)) \wp\, \phi(x) \quad (13)$$

We should note that $\wp(\phi(x) - \phi_0(x))$ is a step function defining the role of how we integrate Eqn 8 above in momentum space via noting that

$$\wp \cdot \phi(x) \equiv \prod_n \frac{1}{L} \cdot \left\{ \frac{d}{d \cdot k_n} \cdot [\exp(-i \cdot k_n \cdot x) \cdot \phi(k_n)] \right\} \cdot d \cdot k_n \tag{14}$$

Which by integration by parts and a momentum based functional derivative we shall define as

$$(D_\psi T)(\phi) = \lim_{\varepsilon \to 0} \frac{T(\phi + \varepsilon \cdot \tilde{\psi}) - T(\phi)}{\varepsilon} \tag{15}$$

leads to if we refer to a Dirac measure at x with $\tilde{\psi} = \delta_x$

$$\frac{\delta}{\delta \phi(x)} T(\phi) \equiv D_{\delta_x} T(\phi) \tag{16}$$

This is due to

$$\delta \cdot T \equiv T(\vec{\phi}(x) + \delta \cdot \vec{\phi}(x)) - T(\vec{\phi}(x)) = \sum_j \left( \frac{\partial T}{\partial \phi_j} \right) \delta \phi_j(x) \equiv \int \left( \frac{\partial T}{\partial \phi} \right) \cdot \delta \phi(x) \cdot dx \tag{17}$$

This in our situation will lead to

$$\int U \frac{\delta V}{\delta \phi} \cdot \wp \phi = UV - \int V \frac{\delta U}{\delta \phi} \cdot \wp \phi \tag{18}$$

As well as

$$\frac{\delta \vartheta(\phi - \phi_0)}{\delta \phi} \cong \delta[\phi - \phi_0] \tag{19}$$

This goes to a tunneling Hamiltonian whose modulus we found was proportional to a S-S' pair 'current'. After a great deal of work, as well as noting we[14] assumed using a scaling of $\hbar \equiv 1$, in which the 'current' becomes[13,14]

$$I \propto \tilde{C}_1 \cdot \left[ \cosh\left[ \sqrt{\frac{2 \cdot E}{E_T \cdot c_V}} - \sqrt{\frac{E_T \cdot c_V}{E}} \right] \right] \cdot \exp\left( -\frac{E_T \cdot c_V}{E} \right) \tag{20}$$

This is due to evaluating our tunneling matrix Hamiltonian with the momentum version of an F.T. of the thin wall approximation, which is alluded to in Fig. 2 [13,14] being set by

$$\phi(k_n) = \sqrt{\frac{2}{\pi}} \cdot \frac{\sin(k_n L / 2)}{k_n} \tag{21}$$

We also assume a normalization of the form[13,14]

$$C_i = \frac{1}{\sqrt{\int_0^{\sqrt{\frac{L^2}{2 \cdot \pi}}} \exp\left( -2 \cdot \{\ \}_i \cdot \phi^2(k) \right) \cdot d\phi(k)}} \tag{22}$$

In doing this, $\{\ \}_i$ refers to initial and final momentum state information of the wave functional integrands obtained by the conversion of our initial and final CDW wave functional states to a $\phi(k)$ 'momentum' basis. Interested readers can get more of the details of this derivation via looking at a mathematical physics arXIV article I wrote which gives more explanations.as to how Eqn (22) was formed from a $\phi(k)$ momentum basis[16]. We evaluate for $i = 1,2$ representing the initial and final wave functional states for CDW transport via the error function

$$\int_0^{\sqrt{\frac{L^2}{2 \cdot \pi}}} \Psi_i^2 \cdot d\phi(k_n) = 1 \tag{23}$$

due to an error function behaving as[13,16,17]

$$\int_0^b \exp(-a \cdot x^2) dx = 1/2) \cdot \sqrt{\frac{\pi}{a}} \cdot erf(b \cdot \sqrt{a}) \tag{24}$$

leading to a renormalization of the form[14]

$$\tilde{C}_1 \equiv \frac{C_1 \cdot C_2}{2 \cdot m_{e^-}} \tag{25}$$

The current expression is a great improvement upon the phenomenological Zener current[1,13,14,18] expression, where $G_P$ is the limiting Charge Density Wave (CDW) conductance.

$$I \propto G_P \cdot (E - E_T) \cdot \exp\left(-\frac{E_T}{E}\right) \text{ if } E > E_T \tag{26}$$

$$0 \qquad otherwise$$

Fig. 5 illustrates to how the pinning gap calculation improve upon a phenomenological curve fitting result used to match experimental data. The most important feature here is that the theoretical equation takes care of the null values before thre threshold is reached by itself. I.e. we do not need to set it to zero as is done arbitrarily in Eqn (26)

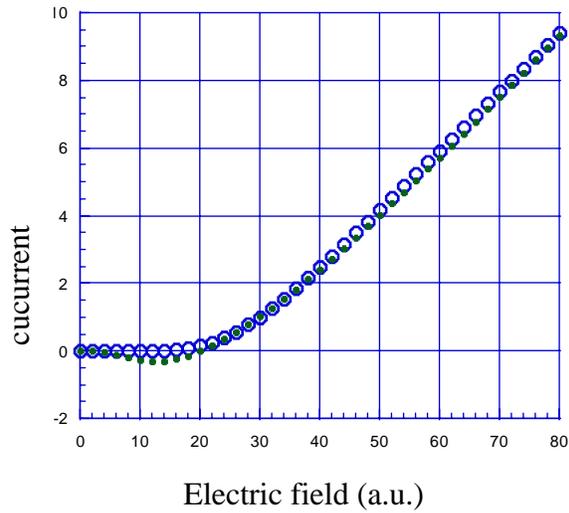

Electric field (a.u.)

Fig 5

Beckwith

*Experimental and theoretical predictions of current values versus applied electric field. The dots represent a Zenier curve fitting polynomial, whereas the blue circles are for the S-S' transport expression derived with a field theoretic version of a tunneling Hamiltonian.*

The Bloch bands are tilted by an applied electric field when we have $E_{DC} \geq E_T$ leading to a S-S' pair shown in Fig. 2.[13,14,19]

The slope of the tilted band structure is given by $e^* \cdot E$ and the separation between the S-S' pair is given by, as referred to in Fig. 2. Note that the $e^* \equiv 2 \cdot e^-$ due to the constituent components of a S-S' pair. And Fig 2 gives us the following distance, $L$, where $\Delta_s$ is a 'vertical' distance between the two band structures tilted by an applied electric field, and $L$ is the distance between the constituent S-S' charge centers.

$$L = \left(\frac{2 \cdot \Delta_s}{e^*}\right) \cdot \frac{1}{E} \qquad (27)$$

So then,[4] we have $L \propto E^{-1}$. When we consider a Zener diagram of CDW electrons with tunneling only happening when $e^* \cdot E \cdot L > \varepsilon_G$ where $e^*$ is the effective charge of each condensed electron and $\varepsilon_G$ being a pinning gap energy, we find .[13,14]

$$\frac{L}{x} \equiv \frac{L}{\bar{x}} \cong c_v \cdot \frac{E_T}{E} \qquad (28)$$

Here, $c_v$ is a proportionality factor included to accommodate the physics of a given spatial (for a CDW chain) harmonic approximation of

$$\bar{x} = \bar{x}_0 \cdot \cos(\omega \cdot t) \Leftrightarrow m_{e^-} \cdot a = -m_{e^-} \cdot \omega^2 \cdot \bar{x} = e^- \cdot E \Leftrightarrow \bar{x} = \frac{e^- \cdot E}{m_{e^-} \omega^2} \quad (29)$$

Realistically, an experimentalist[14] will have to consider that $L \gg \bar{x}$, where $\bar{x}$ an assumed reference point is an observer picks to measure where a S-S' pair is on an assumed one-dimensional chain of impurity sites. The $\bar{x}$ so referred to is given by Eqn (29) above

## IV. COMPARISON WITH LIN'S GENERALIZATION

In a 1999, Qiong-gui Lin[4] proposed a general rule regarding the probability of electron-positron pair creation in D+1 dimensions, with D varying from one to three, leading, in the case of a pure electric field, to

$$w_E = (1 + \delta_{d3}) \cdot \frac{|e \cdot E|^{(D+1)/2}}{(2 \cdot \pi)^D} \cdot \sum_{n=1}^{\infty} \frac{1}{n^{(D+1)/2}} \cdot \exp\left(-\frac{n \cdot \pi \cdot m^2}{|e \cdot E|}\right) \quad (30)$$

When D is set equal to three, we get (after setting $e^2, m \equiv 1$)

$$wIII(E) = \frac{|E|^2}{(4 \cdot \pi^3)} \cdot \sum_{n=1}^{\infty} \frac{1}{n^2} \cdot \exp\left(-\frac{n \cdot \pi}{|E|}\right) \quad (31)$$

which, if graphed gives a comparatively flattened curve compared w.r.t. to what we get when D is set equal to one (after setting $e^2, m \equiv 1$)

$$wI(E) = \frac{|E|^1}{(2 \cdot \pi^1)} \cdot \sum_{n=1}^{\infty} \frac{1}{n^1} \cdot \exp\left(-\frac{n \cdot \pi}{|E|}\right) \equiv -\frac{|E|}{2 \cdot \pi} \cdot \ln\left[1 - \exp\left(-\frac{\pi}{E}\right)\right] \quad (32)$$

which is far more linear in behavior for an e field varying from zero to a small numerical value. We see these two graphs in Fig. 6.

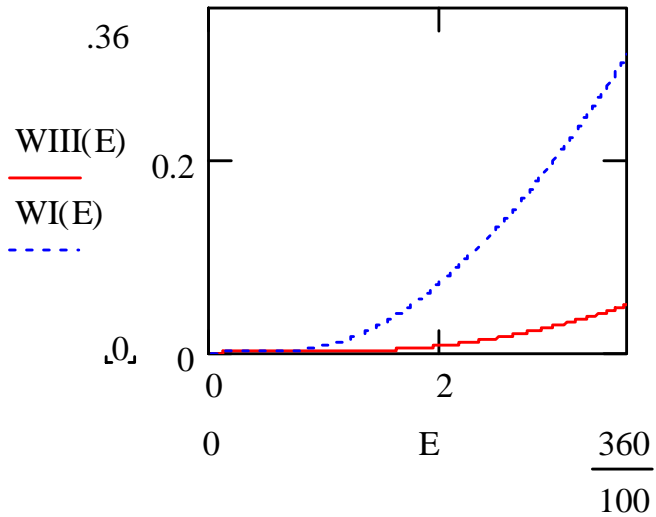

Fig 6

Beckwith

*Two curves representing probabilities of the nucleation of an electron-positron pair in a vacuum. wI(E) is a nearly-linear curve representing a 1+1 dimensional system, whereas the second curve is for a 3 + 1 dimensional physical system and is far less linear*

This is indicating that, as dimensionality drops, we have a steady progression toward linearity. The three-dimensional result given by Lin[4] is merely the Swinger[3] result observed in the 1950s. When I have $D = 1$ and obtain behavior very similar to the analysis completed for the S-S' current argument just presented,[14] the main difference is in a threshold electric field that is cleanly represented by our graphical analysis. This is a major improvement in the prior curve fitting exercised used in 1985 to curve-fit data.[1]

## V. CONCLUSION

We restrict this analysis to ultra fast transitions of CDW;[13,14] this is realistic and in sync with how the wavefunctionals used are formed in part by the fate of the false vacuum hypothesis.

Additionally, we explore the remarkable similarities between what we have presented here and Lin's [4] expansion of Schwinger's[3] physically significant work in electron-positron pair production. That is, the pinning wall interpretation of tunneling for CDW permits construction of *I-E* curves that match experimental data sets; beforehand these were merely Zener curve fitting polynomial constructions.[13,14] Our new physics are and useful for an experimentally based understanding of transport problems in condensed matter physics. Having obtained the *I-E* curve similar to Lin's results[4] gives credence to a pinning gap analysis of CDW transport, with the main difference lying in the new results giving a definitive threshold field effect, whereas both the Zenier curve fit polynomial[1] and Lin's results[4] are not with a specifically delineated threshold electric field. v The derived result does not have the arbitrary zero value cut off specified for current values below given by Miller et al[1] in 1985 but gives this as a result of an analytical derivation.[13,14] This assumes that in such a situation that the electric field is below a given threshold value. In doing this, the final results depend upon a wave functional presentation not materially different from mathematical results derived by Javir Casahoran[20] to represent instanton physics. Furthermore, what is done here is a simpler treatment of transport modeling as is seen in older treatment in the literature[21].

We should note that we used the absorption of a Peierls gap $\Delta'$ term as clearly demonstrated in a numerical simulation paper I wrote.[13] to help form solitons (anti-solitons) used in my Gaussian wave functionals for the reasons stated in my. IJMPB[13] article. This is new physics which deserves serious further investigation. It links our formalism formally with a JJ (Josephon junction) approach, and provides analogies worth pursuing in a laboratory environment. The also stunning development is that the plotting of Eqn (20) ties in with the electron-positron plots as given by Lin in Fig (6)for low dimensional systems, which conveniently fits with identification of a S-S' pair with different 'charge centers' .This tying of formalism with the absorption of a Peierls gap $\Delta'$ term as being necessary for the existence of soliton(anti soliton) 'objects' suggests that the analogies suggested in this paper are now a viable research area waiting to be investigated But we should note that Eqn (20) as a theoretical equation takes care of the null values before thre threshold is reached by itself. I.e. we do not need to set it to zero as is done arbitrarily in Eqn (26). This is a stunning result which in itself is unique.

## VI. ACKNOWLEGEMENTS

The author wishes to thank Dr. John Miller for introducing this problem to him in 2000, as well as for his discussions with regards to the role bosonic states play in affecting the relative power law contribution of the magnitude of the absolute value of the tunneling matrix elements used in the current calculations. In addition, Dr. Leiming Xie highlighted the importance of Eq. (20) as an improvement over Eq. (26) in this problems evaluation. The author claims credit for noticing Lin's derivation as to its relationship to Eq. (20) given in this document.

# REFERENCES

[1] J.H. Miller, J. Richards, R.E. Thorne, W.G. Lyons and J.R. Tucker, *Phys.Rev. Lett. 55*, 1006 (1985)

[2] G. Gruner, *Reviews of Modern Physics*, vol 60, No. 4, October 4, 1988, pp. 1129-1181

[3] J. Schwinger, *Phys.Rev.82*, 664 (1951)

[4] Q.-G. Lin, *J. Phys. G25*, 17 (1999)

[5] E. Tekman, *Phys.Rev.* B 46, 4938 (1992)

[6] J. H. Miller, Jr., C. Ordonez, and E. Prodan, *Phys. Rev. Lett* **84**, 1555 (2000)

[7] J. H. Miller, Jr., G. Cardenas, A. Garcia-Perez, W. More, and A. W. Beckwith, *J. Phys. A: Math. Gen.* **36**, 9209 (2003)

[8] K. Maki; *Phys.Rev.Lett.* **39**, 46 (1977), K. Maki *Phys.Rev* **B 18**, 1641 (1978)

[9] S. Coleman, *Phys.Rev.D 15*, 2929 (1977)

[10] A. Beckwith math-ph/0411031: *An open question: Are topological arguments helpful in setting initial conditions for transport problems in condensed matter physics?* Modern physics letters B, Vol. 20, No 5 (2006), pp 233-243, math-ph/0411031

[11] Private discussions with Dr. J.H. Miller about experimental phenomenology he observed in the device development laboratory, 1998-2000, TcSAM/ U. of Houston

[12] Wen-Fa Lu, Chul Koo Kim, Jay Hyung Lee, and Kyun Nahm, *Phys Rev, D 64*, 025006 (2001).

[13] A. Beckwith, *Making an analogy between a multi chain interaction in charge density wave transport and the use of wave functionals to form S-S' pairs* Int. J. Mod Phys B **19** (24) 2005, pp 3683-3723, arXIV math-ph/0406053

[14] A. Beckwith, PhD dissertation, 2001 (U. of Houston) '*Classical and Quantum models of Density wave Transport: A comparative study*'; E. Kreyszig, *Introductory functional analysis with Applications*, Wiley, 1978, pp 102- 102; Nirmala Prakash, *Mathematical Perspectives on Theoretical physics*, Imperial College Press, 2000, p545-6, in particular 9c.77

[15] R. P. Feyman, R. B. Leighton, and M. Sands, *The Feynman Lectures on Physics, Vol. III*, Addison-Wesley (1964)

[16] A. Beckwith, arXIV math-ph/0406053: *The tunneling Hamiltonian representation of false vaccum decay: II. Application to soliton-anti soliton pair creation*

[17] *CRC Standard Mathematical Tables and Formulas, 30th Edition* CRC Press, (1996), pp. 498-499

[18] H. Miller,J. Richards,R.E. Thorne,W.G.Lyons and J.R. Tucker, *Phys.Rev. Lett. 55*,1006 (1985)

[19] John Bardeen, *Phys.Rev.Lett. 45*, 1978 (1980)

[20] Javir Casahoran, *Comm. Math. Sci*, *Vol 1, No. 2*, pp 245-268.

[21] For a reviews see: R. Jackiw's article in *Field Theory and Particle Physics*, O.; Eboli, M. Gomes, and A. Samtoro, Eds. (World Scientific, Singapore, 1990); F. Cooper and E. Mottola, Phys. Rev. D36, 3114 (1987); S.-Y. Pi and M.; Samiullah, *Phys. Rev.* D36, 3128 (1987); R. Floreanini and R. Jackiw, *Phys. Rev*. D37, 2206 (1988); D. Minic and V. P. Nair, *Int. J. Mod. Phys*. A 11, 2749.